\documentclass[12pt]{iopart}

\usepackage{xcolor}
\usepackage{graphicx}
\usepackage{subcaption}
\usepackage{booktabs}
\usepackage{hyperref}
\usepackage{comment}
\usepackage{soul}

\newcommand{\strev}[1]{}
\newcommand{\rev}[1]{#1}
\newcommand{\revcom}[1]{}

\newcommand{\ud}[1]{\mathrm{d}#1}

\expandafter\let\csname equation*\endcsname\relax
\expandafter\let\csname endequation*\endcsname\relax

\usepackage{amsmath}

\begin{document}

\title{Exploring Particle Geodesics in a Warp Drive Spacetime}

\author{Lucas Timotheo Sanches\footnote{Corresponding author}}
\address{Center for Computation and Technology -- Louisiana State University, 888 S. Stadium Dr., Baton Rouge, LA, USA}
\ead{lsanches@lsu.edu}
\vspace{10pt}

\author{Max Morris}
\address{Center for Computation and Technology -- Louisiana State University, 888 S. Stadium Dr., Baton Rouge, LA, USA}
\ead{mmorris@cct.lsu.edu}
\vspace{10pt}

\author{Steven R. Brandt}
\address{Center for Computation and Technology -- Louisiana State University, 888 S. Stadium Dr., Baton Rouge, LA, USA}
\ead{sbrandt@cct.lsu.edu}
\vspace{10pt}

\begin{indented}
\item[]June 2025
\end{indented}

\begin{abstract}
Although the Alcubierre Warp Drive~\cite{Alcubierre_1994} is theoretically capable of providing faster-than-light travel, it may be difficult to use for this purpose. But is it useful for slower-than-light travel? We begin by observing the that the warp bubble will act to protect the ship from dust particles and other space debris (a potentially serious hazard even at 10\% the speed of light). We then explore several modifications of the Alcubierre Warp Drive, e.g. a ``deflector shield,'' with the perspective of keeping a ship safe from collisions with particles, projectiles, rogue planets, and other dangers of space travel.
\end{abstract}

%
\vspace{2pc}

\section{Introduction}\label{sec:intro}

The Alcubierre Warp Derive~\cite{Alcubierre_1994} has provided a fertile ground for many \textit{Gedankenexperiments} exploring the limits of what is possible in a universe governed by Einstein's theory of General Relativity. It also provides a rich playground for studying the behavior of both null and time-like geodesics, as is done in Refs.~\cite{Clark_1999, muller2012detailed}, respectively. This paper will complement these studies by looking at the behavior of massive particles as a model for interstellar debris such as dust, meteors, and other hazards, focusing on the dangers of collisions with these objects \rev{when using sub-luminal warp bubbles}.

Can a ship traveling by \rev{sub-luminal} warp drive collide with particles in its path? If so, at what relative speed will such collisions take place and what can be done to mitigate the danger?

As a safety benchmark, we imagine that such collisions must occur at 10\% the speed of light (as measured relative to the ship) or less. At that speed, even a particle with a mass of $10^{-9} \text{kg}$, a smallish dust particle, will strike with the same kinetic energy as a bowling ball ($\approx 6 \text{kg}$)
traveling at mach $1$. At a mass of $10^{-6} \text{kg}$, it is more like being hit by a
small truck at that same speed. 

We will show that the geodesics of particles in the traditional Alcubierre metric
tend to be slowed or deflected by the warp bubble,
providing modest protection from dust, meteors, etc. We will then go on to investigate mechanisms to improve the safety of the ship.

But why travel slower than light\strev{ if an Alcubierre Drive is possible}?
Firstly, we note that attempts to simulate a superluminal Alcubierre
Drive have met with instability~\cite{clough2024no}. The authors did not
determine the cause for this problem, but instead chose to restrict
their evolutions to slower-than-light systems.

Secondly, we observe that an Alcubierre Drive cannot allow a ship to freely explore
space in any direction at faster-than-light speeds because of a causal disconnect
between the ship and the warp bubble. To see why this is so, consider the Alcubierre Warp Drive Metric\cite{Alcubierre_1994}
\begin{equation}
    ds^2 = -dt^2+dz^2+dy^2+\left(dx - f v dt\right)^2,
\end{equation}
where
\begin{align}
    r & = \sqrt{(x-vt)^2 + y^2 + z^2} ,\label{eq:alcubierre_r} \\
    f(r) & = \frac{\tanh\left[\sigma(r+R)\right] - \tanh\left[\sigma(r - R)\right]}{2\tanh(\sigma R)}, \label{eq:alcubierre_f}
\end{align}
with $v$ representing the coordinate speed of a warp bubble moving along the $x$ direction with radius $R$. The function $f(r)$ is referred to as \textit{form function} by Alcubierre and Lobo in Ref.~\cite{alcubierre2017warp}. This function asymptotically evaluates to $1$ inside the bubble and $0$ at infinity, becoming a ``top-hat'' when $\sigma \rightarrow \infty$, as stated in Ref.~\cite{Alcubierre_1994}.

For this metric, a null geodesic traveling in the $+x$ direction has speed $dx/dt = 1 + f v$.
It is noted in Ref.~\cite{lobo2017wormholes} that when this quantity becomes equal to the bubble speed, $v$, i.e. when $f = 1-1/v$, then it is impossible for light rays to continue to propagate forward. Thus, there is a horizon and a causal disconnect between the ship and the outer layers of the bubble.
Perhaps more disconcerting is a fact that is not pointed out in Ref.~\cite{lobo2017wormholes}, that beyond this point, in the outer portions of the bubble where $f < 1 - 1/v$, the negative mass energy matter that forms the bubble is actually \textit{superluminal}, i.e. $v > dx/dt = 1 + f v$.

Perhaps a more serious blow to the feasibility of the warp drive is that a quantum back-reaction would destabilize any
warp drive that attempts to exceed light speed~\cite{hiscock1997quantum,gonzalez2000warp,barcelo2010impossibility,alcubierre2017warp}.

Finally, there is the problem that any faster-than-light
travel (including the Alcubierre Drive~\cite{everett1996warp}) can lead to closed,
timelike curves and run into causality issues.

On the other hand, for slower-than-light travel, an Alcubierre Drive
can simply be generated by the ship and \textit{engaged} to travel in any direction.

We also consider the possibility of encountering a rogue planet. At 10\%
light speed, detecting it in time to turn out of the path of such a hazard
may prove nearly impossible.

However, it turns out that with relatively simple modifications, the Alcubierre
Drive can make slower-than-light travel safe from dust particles, meteors, and
from rogue planets. The main contributions of this paper are (1) the analysis of the potential hazards (represented by time-like geodesics of massive particles)
of near light-speed travel using an Alcubierre Drive, (2) modifications to the warp drive to mitigate such hazards, and (3) the introduction of a new, open-source tool for interactively exploring and identifying interesting particle dynamics in these spacetimes (See Ref.~\cite{core_repo,sim_repo}).

\section{Generic warp drives of the Natário class and its geodesic equations}\label{sec:metric_and_geodesics}

The spacetimes we will consider in this work, including the original Alcubierre~\cite{Alcubierre_1994} solution, fall into the broad category of warp drive spacetimes of the \textit{Natário} class, first introduced in Ref.~\cite{Natario_2002} and studied in detail in Ref.~\cite{santiago_2022}. We will begin with the generic spacetime metric presented in Eq. (2.2) of Ref.~\cite{santiago_2022}. Given the set of Cartesian-like coordinates $q^\mu = (t, x, y, z)$, the general warp drive metric we will consider is given by 
\begin{equation}
    \mathrm{g}_{\mu \nu} = \left(
    \begin{array}{c|c}
        v^2 - 1 & -v_{i} \\
        \hline
        -v_i & \delta_{ij}
    \end{array}
    \right),
    \quad
    \mathrm{g}^{\mu \nu} = \left(
    \begin{array}{c|c}
        -1 & -v^{i} \\
        \hline
        -v^{i} & \delta^{ij} - v^i v^j
    \end{array}
    \right).
    \label{eq:generic_warp_drive_metric}
\end{equation}
We adopt the convention that Greek indices run over all 4 spacetime coordinates, whereas Latin indices run over spatial coordinates $(x, y, z)$. The symbol $\delta_{ij}$ is the Kronecker delta, $v^i = v^i(t, x, y, z)$ is known as the \textit{flow vector} and $v^2 \equiv \left(v^x\right)^2 + \left(v^y\right)^2 + \left(v^z\right)^2$.

Equation~\eqref{eq:generic_warp_drive_metric} is nothing more than the ADM decomposition of a flat spacetime metric with nontrivial shift, given by $-v^i$. It is common to refer to $v^i$ as the \textit{flow vector} in the context of warp drive spacetimes (as opposed to referring to the shift vector of the ADM decomposition).

An important time-like observer arises in ADM decompositions, known as the \textit{Eulerian} observer. This is a ``co-moving'' (see Sec.~2 of Ref.~\cite{santiago_2022}) observer that is orthogonal to the spatial slices of the foliation. In spacetimes with unit lapse, such as those we are considering, its components are given by
\begin{equation}
    n_\mu = \left( -1, 0, 0, 0 \right), \quad n^\mu = \left(1, v^x, v^y, v^z \right).
    \label{eq:eulerian_observer}
\end{equation}
As we shall see later on, this observer represents a static particle solution of the geodesic equation and can be used to infer the behavior of a warp bubble interacting with a field of particle-like debris.

Our work hinges on the study of the trajectories of free-falling particles in the spacetime described by Eq.~\eqref{eq:generic_warp_drive_metric}. Given that the class of generic warp drives metrics we consider are given in ADM form and that we wish to study how dynamically changing spacetime parameters affect the particle dynamics, we will adopt the $3+1$ decomposition of the geodesic equations presented in Eqs. (28b) and (25) of Ref.~\cite{Vincent_2012}. In this formulation, a particle's state is defined by 3 positions, $X^i(t)$, 3 velocities, $V^i(t)$ and an energy, $E(t)$. The particle's velocities and energy are measured by the Eulerian observer of the ADM formulation.

When specifying particle initial data, care must be taken so that the contraction of the 4-velocity, $u^\mu$, with itself, namely
\begin{equation}
    \mathrm{g}_{\mu \nu} u^\mu u^\nu = -\eta,
    \label{eq:norm_momenta}
\end{equation}
is satisfied for all particles, where $\eta = 1$ for massive particles, and $\eta = 0$ for photons.

By utilizing the definition of $p^\mu$, given in Eq.~(8) of Ref.~\cite{Vincent_2012}\rev{,
\begin{equation}
    p^\mu = E (n^\mu + V^\mu),
    \label{eq:vincent_8}
\end{equation}}
together with Eqs.~\eqref{eq:generic_warp_drive_metric} and \eqref{eq:eulerian_observer} and setting $V^{t} = 0$, we can write
\begin{equation}
    E^2\left(1 - \delta_{ij}V^i V^j\right) = \eta,
    \label{eq:norm_square}
\end{equation}
where $E = u_\mu n^\mu = p_\mu n^\mu/m$ is the energy per unity mass.

To initialize a massive particle, we choose arbitrary initial positions and initial 3-velocities such that $\delta_{ij}V^i V^j < 1$. We then solve Eq.~\eqref{eq:norm_square} for $E$ to find
\begin{equation}
    E = \left(\frac{1}{1-\delta_{ij}V^iV^j}\right)^{1/2}.
    \label{eq:norm_condition_1}
\end{equation}
To initialize a photon, an arbitrary initial energy $E$ can be chosen, but we then impose that $\delta_{ij}V^iV^j = 1$, thus satisfying Eq.~\eqref{eq:norm_square}.

A final ingredient common to warp drive spacetimes is a choice of \textit{form function}.
As previously stated, this is an arbitrary function that evaluates to $1$ in the interior of a warp bubble and $0$ on the exterior, with a transition region in between. In his original work, Alcubierre chose Eq.~\eqref{eq:alcubierre_f} for the form function.
In our work, the form
function will be constructed as follows: Let $p(s)$ be a 7th degree polynomial in $s$ given by
\begin{equation}
    p(s) = \left(\Delta_x - s\right)^4 \left( \Delta_x^3 + 4 \Delta_x^2 s +10 \Delta_x s^2 + 20 s^3\right)\Delta_x^{-7}.
    \label{eq:trans_poly}
\end{equation}
Then, the function
\begin{align}
    \theta(x) =
    \begin{cases}
        1, & \text{if $x < x_0$} \\
        0, & \text{if $x > x_0 + \Delta_x$} \\
        p(x - x_0) & \text{otherwise}
    \end{cases}
    ,
    \label{eq:def_f}
\end{align}
smoothly transitions from $\theta = 1$ at $x = x_0$ to $\theta = 0$ at $x = x_0 + \Delta_x$. We refer to $x_0$ as the \textit{transition radius} and $\Delta_x$ as the \textit{transition width}. For smoothness of particle motion and consistency with the Israel-Lanczos-Sen junction conditions (see Sec.~2 of Ref.~\cite{santiago_2022} and Sec. 3.7 of Ref.~\cite{Poisson2007} for further discussion), we have constructed $p(s)$ so that it is $C^3$. We did this by using a 7th order polynomial and requiring that its first, second and third derivatives are zero at $x=x_0$ and $x=x_0 + \Delta_x$.  Numerically, however, the code does not require this and works even with the piecewise linear function $\theta_a$ (see Eq.~\eqref{eq:def_fa}), even though it is $C^0$. Later on, we will use $\theta_a$ in mathematical analyses because it is simpler to work with.
\begin{align}
    \theta_a(x) =
    \begin{cases}
        1, & \text{if $x < x_0$} \\
        0, & \text{if $x > x_0 + \Delta_x$} \\
        1-(x-x_0)/\sigma & \text{otherwise}
    \end{cases}
    .
    \label{eq:def_fa}
\end{align}

The form function given by Eq.~\eqref{eq:def_fa} has clearly discontinuous derivatives across the hypersurfaces, giving rise to matter shells. These shells will be located precisely on the warp bubble interface and on the region where the warp metric transitions back to flat spacetime. Note, however, that we don't need to calculate the distribution of matter to study geodesics, because the motion of particles results only from the gravitational field, i.e. the metric. Once the spacetime metric and its derivatives are is specified, even when they are discontinuous, particle motion is completely determined. Computing the matter configuration that makes this motion possible is a separate calculation.

\section{The Alcubierre Drive as Shield}\label{sec:alcubierre_drive}

In this section, we will study the behavior of massive particles in the Alcubierre warp drive spacetime, but with a different form function: Eq.~\ref{eq:def_f}. This localizes the warp bubble to a compact region, a modification which proved to be useful in our numerical simulations and visualizations. Referring to the definitions of Sec.~\ref{sec:metric_and_geodesics}, the Alcubierre drive is defined by choosing~\cite{Alcubierre_1994,santiago_2022}
\begin{align}
    v^x & = u \, \theta(r;\;R,\,\sigma) \label{eq:alcubierre_vx},\\
    v^y & = 0 \label{eq:alcubierre_vy},\\
    v^z & = 0 \label{eq:alcubierre_vz},
\end{align}
where $u$ is the bubble's constant speed, $R$ is the bubble radius, $\sigma$ the bubble's width, $\theta(r;\;R,\,\sigma) $ is given by Eq.~\eqref{eq:def_f}, and, $r$ measures the distance to the center of a warp bubble with constant velocity $u$ at a given coordinate time $t$, for which we adopt the original prescription of Alcubierre, given by
\begin{equation}
    r = \sqrt{(x - ut)^2 + y^2 + z^2}.
    \label{eq:def_r}
\end{equation}
For this metric, the components of the 4-velocity of the Eulerian observer are given by
\begin{equation}
    n_\mu = \left(-1, 0, 0, 0\right),\quad n^\mu =  \left(1, v^x, 0, 0\right).
    \label{eq:eulerian_obs_alc}
\end{equation}
We can use the Eulerian observer to better understand the behavior of massive, initially static, particles in the vicinity of a warp drive. Given that these observers follow geodesics, we can write
\begin{equation}
    \frac{\ud{q^\mu}}{\ud{\lambda}} = \left(1, v^x, 0, 0\right),
    \label{eq:geodesic_static_particle}
\end{equation}
where $q^\mu(\lambda)$ are the particle's 4-positions parametrized by some affine parameter $\lambda$.

From Eq.~\eqref{eq:geodesic_static_particle}, it is easy to see that particles outside the warp bubble will experience no change in their positions, since in this region $\theta(r; R, \sigma) \rightarrow 0$. Inside the bubble, however, particles are ``dragged along'' with velocity $u$, since in this region $\theta(r; R, \sigma) \rightarrow 1$. With this, we can infer what happens when an Alcubierre Drive encounters a stationary debris field, which we will model as an ensemble of static particles. 

Initially, when the bubble is far away from a debris field, the particles experience no change in their positions. As the bubble gets closer, particles slowly acquire the 4-velocity $u$, as $\theta(r; R, \sigma)$ transitions to $1$. They will then be ``stuck'' there, moving with the bubble. This means that the Alcubierre Drive effectively ``picks up'' debris (or static particles, in this case) as it travels through space.

\subsection{Numerical Simulations and trajectory visualizations}

To simplify the discussion and analysis of this paper, we will consider warp drives that travel at one half the speed of light only. Debris fields to be considered will primarily be particles with zero initial velocity, or particles moving at $1\%$ the speed of light in the warp drive coordinate system. This speed is both small enough to be a perturbation (from a mathematical viewpoint), and unreasonably fast (from a physical viewpoint).

We adopt units such that $c=G=1$ and the mass of the sun is $m_\odot=1$. This means a unit length is $1.47713754$ km and a unit of time is $4.92720047$ $\mu\text{s}$, but we stress that the physics of this system should work equally well with any choice of unit mass. The inner part of the warp bubble will be placed at a distance of $R=4$ from the ship, and the radius of the bubble will be $\sigma=4$, approximately 6 km. In these units, the sort of dust particles we are considering will have masses in the range $10^{-37}$ to $10^{-40}$, and the mass of the ship (assuming 200 metric tons), will be about $10^{-22}$.

In all trajectory plots, the ship is represented by a freely available, copyright-free image, found in Ref.~\cite{spaceships}. The warp bubble radius is drawn as a pair of black, dashed circles around the ship. They are the results of a contour plot with lines drawn at $1\%$ and $99\%$ warp field strength. The locations are, approximately, $R$ and $R+\sigma$. When a deflector shield is present (to be described later), it will be drawn with a dotted line marking the contour of $10\%$ strength. Trajectories of particles will be drawn with dot-dashed lines.

Trajectories of particles are computed using a pair of highly configurable \texttt{Rust}-based codes. One code implements the basic physics of the simulation (which we generated in Mathematica verified using our \texttt{EinsteinEngine} framework~\cite{einstein-engine}) and can be found in Ref.~\cite{core_repo}. The other code implements a real time visualizer tool and can be found in Ref.~\cite{sim_repo}. Using this system, a wide variety of physical systems may be modeled, monitored, and explored. The code can evolve both massive and photonic particles, or a mixture of those types. Initial data can be constructed, or the results of a simulation analyzed using a \texttt{Python} framework (included in the visualizer repository) that understands \texttt{JSON} data files produced by the visualizer.

In order to simulate particle trajectories in real time, it is convenient to advance the state of the particles by a fixed time step. To do so, we employ the \texttt{RK4} algorithm with a configurable time step that defaults to $dt=0.2$. For each particle, and at every time step, we check that $\eta$ is either $1$ or $0$ to within a configurable tolerance which defaults to $10^{-6}$. If it is not, we take two time steps of size $\overline{dt}=dt/2$. We found that this ``adaptiveness'' mechanism greatly improves particle normalization conservation and overall numerical behavior. It is well known that geodesic systems can be numerically ``stiff,'' which justifies the need of the special treatment just described. As an additional validation mechanism, we check for \texttt{NaNs} in the state vectors of all particles.

\subsection{The Non-Relativistic Debris Field}

To illustrate the behavior of interaction with stationary particles, we evolve the system from an initial vertical line of particles, then we select a representative handful for which to draw trajectories.

\begin{figure}[ht]
    \centering
    \begin{subfigure}{0.33\textwidth}
    \includegraphics[width=\linewidth]{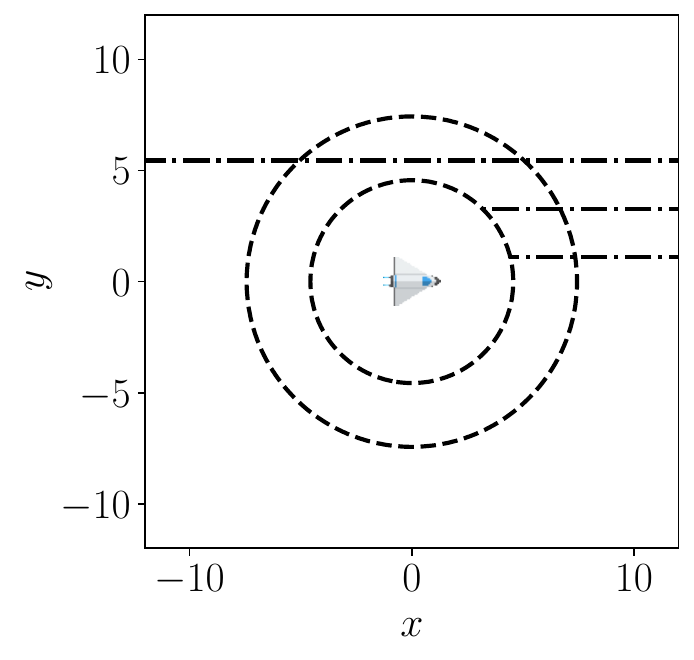}
    \caption{}\label{fig:plain_warp}
    \end{subfigure}%
    \begin{subfigure}{0.33\textwidth}
    \includegraphics[width=\linewidth]{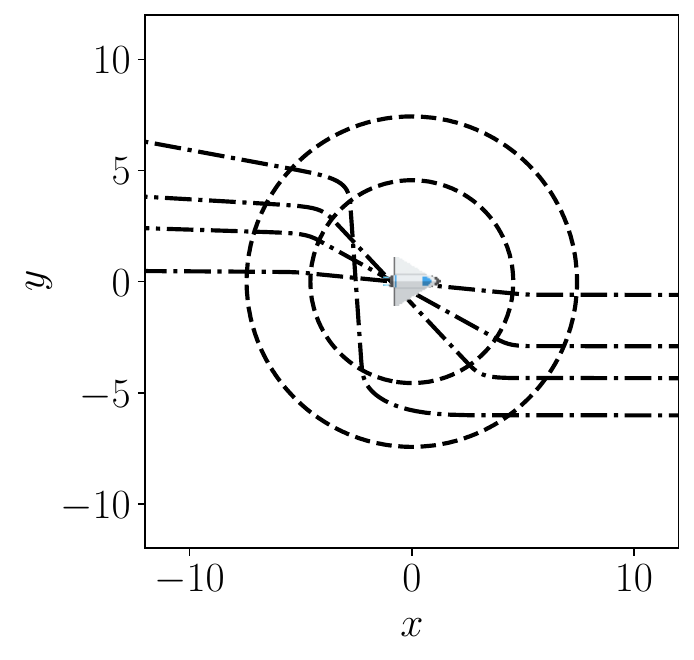}
    \caption{}\label{fig:plain_warp_left}
    \end{subfigure}%
    \begin{subfigure}{0.33\textwidth}
    \includegraphics[width=\linewidth]{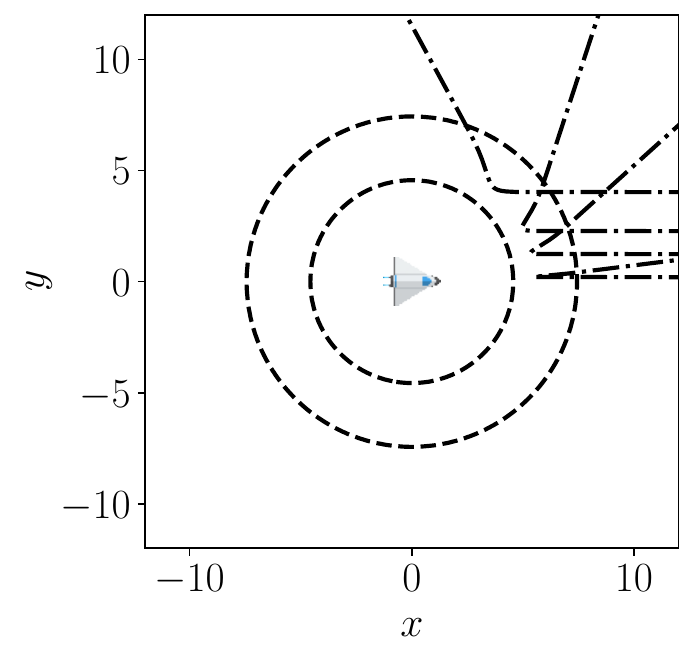}
    \caption{}\label{fig:plain_warp_right}
    \end{subfigure}%
    \caption{
      Fig~\ref{fig:plain_warp}: a warp drive ship moving to the right (positive $x$) through a field of stationary particles. The two dashed concentric circles represent the inner and outer radii of the warp field. Any particle that impacts on the inner edge of the bubble comes to a stop relative to the ship. Fig~\ref{fig:plain_warp_left} a warp drive moving through a field of particles that are moving to the left at $1\%$ light speed. The particles move through the warp field and are focused by it. Inside the bubble, particle speeds reach $10\%$ the speed of light. Fig.~\ref{fig:plain_warp_right}: a warp drive moving through a field of particles moving to the right at $1\%$ light speed. The particles are reflected by the warp field and reach $80\%$ light speed.
    }
    \label{fig:plain_warp_all}
\end{figure}

In addition to the debris field, we also evolve a static particle placed at the origin. This particle models the trajectory of the ship inside the warp bubble. Evolving it serves as a correctness check of our code. We plot the results of the evolution in Fig.~\ref{fig:plain_warp}. Particles that are outside the central part of the bubble travel straight through, eventually escaping the warp field. Particles intersecting the central part of the bubble come to a stop on its surface.

Thus, a warp drive seems to protect a ship from collision with stationary particles.

\subsubsection{Analysis for Stationary Particles}
The above results can be derived analytically from the geodesic equations. To do this, first consider a massive particle with velocity vector
\begin{equation}
    V^i = \left[ V^x(t), 0, 0 \right],
    \label{eq:small_v_vector}
\end{equation}
and corresponding 4-velocity
\begin{equation}
    u^\mu = \left[1,\, V^x(t) + u \theta(r;\; R,\,\sigma),\, 0,\, 0 \right].
    \label{eq:small_p_vector}
\end{equation}
With this choice, the non-zero components of the $3+1$ geodesic equations become
\begin{align}
    \frac{\ud{X}}{\ud{t}} & = V^x + u \, \theta(r; R, \sigma),    \label{eq:small_v_start1}\\
    \frac{\ud{V^x}}{\ud{t}} & = -u V^x \left[ 1 - \left(V^x\right)^2\right] \theta^\prime(r; R, \sigma). \label{eq:small_v_end1}
\end{align}

Eqs.~\eqref{eq:small_v_start1}--\eqref{eq:small_v_end1} can be solved analytically under some simplifying assumptions. Firstly, we replace $\theta(r; R,\sigma)$ by the linear form function $\theta_a(r; R, \sigma)$ given by Eq.~\eqref{eq:def_fa}.
Secondly, we will constrain ourselves to analyzing a particle while it is on the $x$-axis and in the transition region, and therefore
\begin{equation}
    \theta_a(r; R, \sigma) = \frac{R + \sigma - r}{\sigma}.
\end{equation}
Given a particle that is initially at rest, i.e. $V^x(0)=0$, the solution to Eq.~\eqref{eq:small_v_end1} is that $V^x(t)$ will remain zero. The solution for $X(t)$ on the other hand, is given by
\begin{equation}
    X(t) = R+\sigma  e^{-\frac{t u}{\sigma }}+u t.
    \label{eq:v0_is_zero}
\end{equation}
To find the location of the particle relative to the ship, one just needs to subtract off $ut$ from Eq.~\eqref{eq:v0_is_zero}. It is then clear that the particle exponentially converges to the inner radius of the bubble. Even though Eq.~\eqref{eq:v0_is_zero} was obtained for the special case of a particle on the $x$ axis, we have numerically observed that the result holds true for out-of-axis particles in our simulator code.

\subsection{The Relativistic Debris Field}

A more interesting scenario occurs when particles are allowed to have non-zero initial velocities. To investigate this, we make use of a wall of particles positioned to the right of the ship with a velocity of $\pm 1\%$ of light speed. 

\subsubsection{Negative Initial Velocity}

When space debris have negative velocities and move to the left (see Fig.~\ref{fig:plain_warp_left}) they penetrate the bubble and are focused on a region just behind the ship. During this process they are also accelerated, reaching $10\%$ of light speed inside the bubble near the ship. This presents a potential hazard. We note, however, that it is a smaller hazard than colliding with a particle traveling at $50\%$ light speed, and so the warp bubble still offers some protective effect.

During the course of our experiments, we also noticed that when the initial particle velocity is small, its speed inside the bubble is $\sqrt{v_0}$. Thus, e.g. an initial speed of $v_0=1.0\times 10^{-4}$ results in a speed of $v_0=1.0\times 10^{-2}$ inside the bubble.

\subsubsection{Analysis for Negative Initial Velocity}

Given an initial velocity of $v_0$, we found two solutions to Eq.~\eqref{eq:small_v_start1}--~\eqref{eq:small_v_end1}. The following set of equations
should be used when $v_0 < 0$, i.e. it is moving to the left and toward the ship:
\begin{align}
    \Xi & = \sqrt{e^{\frac{2 t u}{\sigma }}+1/v_0^2-1},\label{eq:sol1_start} \\
    V^x & =-e^{\frac{t u}{\sigma }}/\sqrt{\Xi },\label{eq:sol1_vx}\\
    x &= e^{-\frac{t u}{\sigma }} \left(v_0 \left(u (r+t u) e^{\frac{t u}{\sigma }}+\sigma  \left(u-\sqrt{\Xi }\right)\right)-\sigma \right)/(u v_0).
    \label{eq:sol1}
\end{align}
In the above, the quantity $\Xi$ is factored out only to simplify the expressions.

As the particle travels through the transition region, it also speeds up, see Eq.~\eqref{eq:sol1_vx}. This acceleration increases until it reaches the inner radius of the bubble. Using the above solution for $x$, see Eq.~\eqref{eq:sol1}, one can solve for the time and the final velocity. At this point, the particle's velocity is
\begin{equation}
    v_f = \frac{\sqrt{-v_0 \left(\left(u^2+1\right) \left(-v_0\right)-2 u\right)}}{u \left(-v_0\right)-1}.
\end{equation}
In the limit of small $v_0$, this reduces to
\begin{equation}
    v_f = -\sqrt{2} \sqrt{u} \sqrt{-v_0}+O\left(\left(-v_0\right){}^{3/2}\right).
\end{equation}
This agrees with our observation that for $u=1/2$, particles with small inward velocities are accelerated to a speed close to $\sqrt{v_0}$. We have tested with bubble speeds other than $u=1/2$ and found that the formula holds.

\subsubsection{Positive initial velocity}
When space debris have positive velocities and move to the right (see Fig.~\ref{fig:plain_warp_right}), particles are reflected from the bubble and attain final speeds as much as $80\%$ of light speed. To conserve energy and momentum of the system, there should be a tiny change in the momentum and energy of the ship. Although we are assuming that for any individual particle this is negligible, the collective action of many dust particles accelerated to high speeds would probably serve to slow the ship.

Both the reflection and focusing effects of the warp drive remain similar, even when the speed of particles is as small as $10^{-6} c$.

The consequence of our analysis for negative and positive $v_0$ is that the analytic result that we obtained for stationary particles (Eq.~\eqref{eq:v0_is_zero}) is unlikely to ever occur. Any small deviation from $v_0=0$ will result in acceleration away from that point.

\subsubsection{Analysis for Positive Velocity}

As stated previously, given an initial velocity of $v_0$, we found two solutions to equations Eq.~\eqref{eq:small_v_start1}--\eqref{eq:small_v_end1}. The set below should be used when $v_0 > 0$, i.e., it is moving to the right and away from the ship:
\begin{align}
    \Xi & = \sqrt{e^{\frac{2 t u}{\sigma }}+1/v_0^2-1}, \\
    V^x & =e^{\frac{t u}{\sigma }}/\Xi,\label{eq:pos_v0_vx}\\
    x &= e^{-\frac{t u}{\sigma }} \left(v_0 \left(u (r+t u) e^{\frac{t u}{\sigma }}+\sigma  (u+\Xi )\right)-\sigma \right)/(u v_0).\label{eq:pos_v0_x}
\end{align}

\begin{figure}
    \centering
    \includegraphics[width=0.5\linewidth]{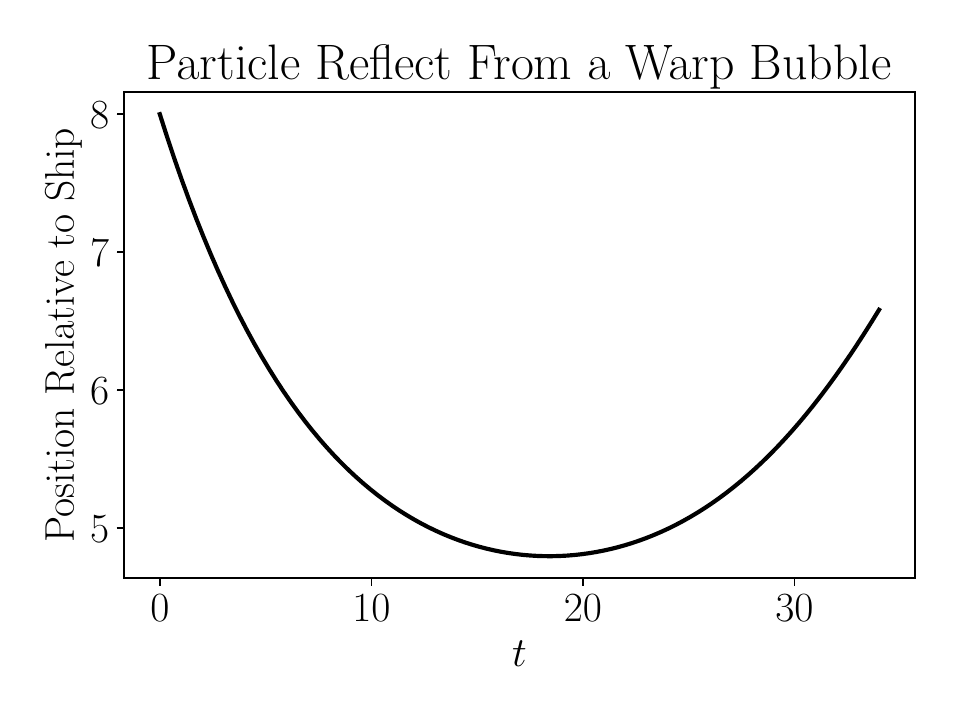}
    \caption{Trajectory of a particle with a small initial velocity away from the ship while inside the transition region of a warp drive.}
    \label{fig:reflect}
\end{figure}

The trajectory of a particle with $v_0=0.01$ is shown in Fig.~\ref{fig:reflect}. The result is a particle that is reflected while its speed grows rapidly. One can compute the particle's final speed by determining when Eq.~\eqref{eq:pos_v0_x} is equal to $R+\sigma$ and substituting back into Eq.~\eqref{eq:pos_v0_vx}:
\begin{equation}
    v_f = \frac{\left(u^2+1\right) v_0-2 u}{\left(u^2-1\right) v_0 \sqrt{\frac{\left(u^2-2 u v_0+1\right){}^2}{\left(u^2-1\right)^2 v_0^2}}}.
\end{equation}
In the limit that $v_0$ is small, we have
\begin{equation}
    v_f = \frac{2 u}{u^2+1}-\frac{\left(u^2-1\right)^2 v_0}{\left(u^2+1\right)^2}+O\left(v_0^2\right).\label{eq:vf_pos_small_v0}
\end{equation}
The first two terms in Eq.~\eqref{eq:vf_pos_small_v0} are what you would expect from a special relativistic particle reflecting from a hard surface. The result agrees with our numerical simulation, that for the configuration we are studying, particles moving slowly to the right will be reflected at about $80\%$ the speed of light. Note that this result does not depend on the radius or thickness of the bubble, only the speed of the ship.

\begin{figure}[ht]
    \centering
    \begin{subfigure}{0.33\textwidth}
    \includegraphics[width=\linewidth]{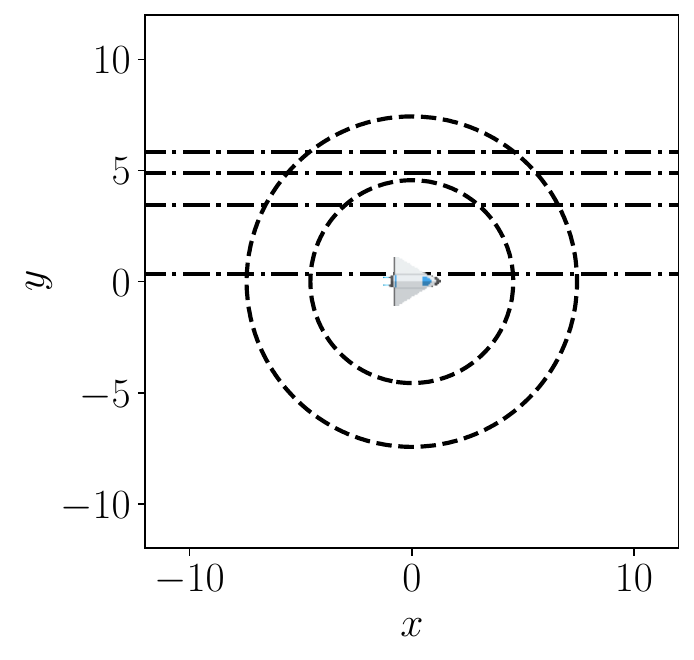}
    \caption{}\label{fig:positive_slippage}
    \end{subfigure}%
    \begin{subfigure}{0.33\textwidth}
    \includegraphics[width=\linewidth]{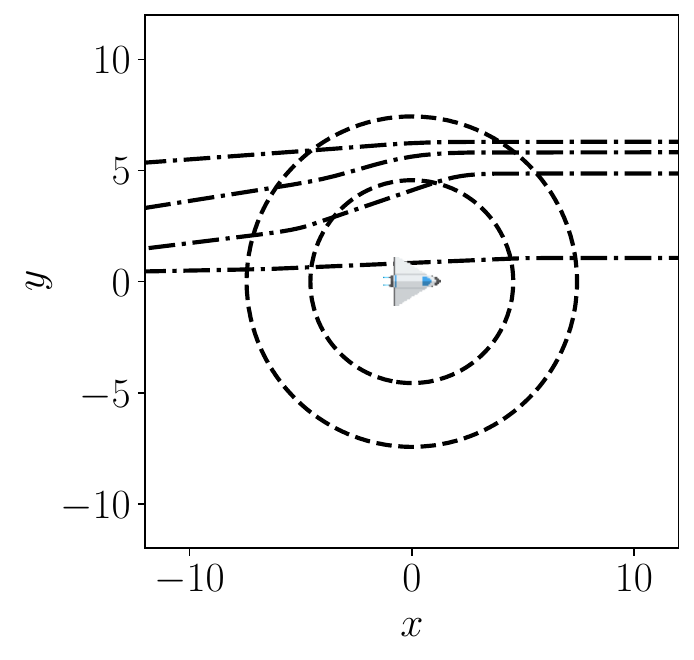}
    \caption{}\label{fig:positive_slippage_left}
    \end{subfigure}%
    \begin{subfigure}{0.33\textwidth}
    \includegraphics[width=\linewidth]{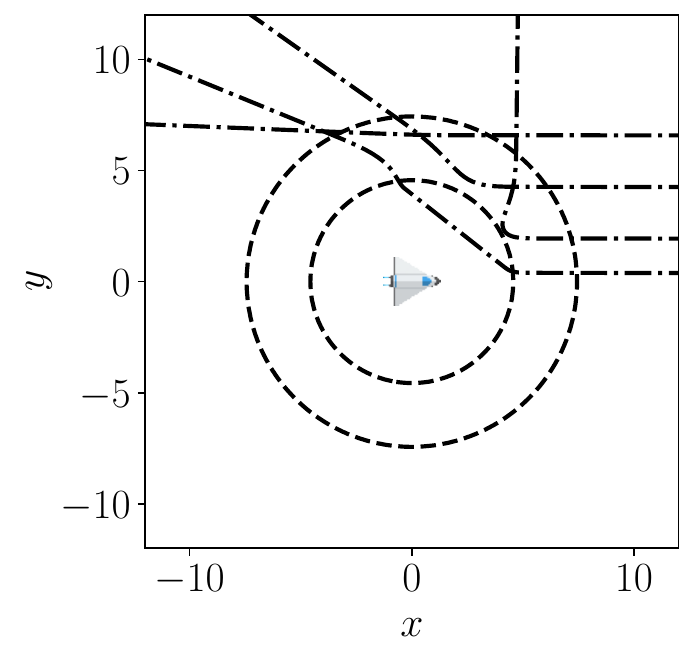}
    \caption{}\label{fig:positive_slippage_right}
    \end{subfigure}%
    \caption{
      Fig~\ref{fig:positive_slippage}: a warp drive moving to the right (positive $x$) through a field of stationary particles with positive slippage. The two dashed concentric circles represent the inner and outer radii of the warp field. The particles are able to penetrate the shield and go straight through. Fig~\ref{fig:positive_slippage_left} shows particles traveling to the left with positive slippage. The behavior is similar to the stationary particles. Fig.~\ref{fig:positive_slippage_right}: a warp drive moving through a field of particles moving to the right at $1\%$ light speed. The particles are reflected, but much more gently than without slippage.
    }
    \label{fig:positive_slippage_all}
\end{figure}

\section{Slippage: Distinct Bubble and Interior Speeds}\label{sec:slippage}

One of the problems with the warp drive discussed in the previous section was the
high speed of reflected particles. We hypothesized that this might result in a large expenditure of energy by the ship.

One way to mitigate this energy drain is through a concept we will call
\textit{slippage}. Instead of having the ship be at rest with respect to the bubble's frame, we will give it some
positive velocity.
This means that only part of the ship's forward progress will result from the frame-dragging induced by the warp bubble, while the rest will be produced by its own engines.

To represent this mathematically, we modify $v_x$ such that
\begin{equation}
    v^x = u_d \, \theta(r;\; R,\, \sigma)
    \label{eq:slippage},
\end{equation}
where $u_d$ is the magnitude of the shift inside the bubble (which we refer to as the \textit{drag speed}), and $u_b$ is the speed at which the bubble moves through coordinate space.

The 4-velocity of the ship is now represented by
\begin{equation}
    u_\mu = \frac{1}{\sqrt{1-u_s^2}}\left(-1,u_s,0,0\right),\label{ship_momentum}
\end{equation}
where $u_s = u_b - u_d$ is the ship speed.

We note that Eq.~\eqref{ship_momentum} describes a particle
moving in coordinate space with speed
$u_b$, if and only if it is inside the bubble.

In this scenario, if a particle that is initially at rest collides with the bubble, by the time it
reaches the interior, it will still have a positive speed, and so it will continue
to travel to the left, moving at a speed relative to the ship. We call this a \textit{positive slippage}.

To illustrate this concept, we once again study the same three debris fields. See Fig.~\ref{fig:positive_slippage_all}. When comparing these results to those of Fig.~\ref{fig:plain_warp_all}, we can see that the debris does not get reflected to the same degree. The particles that were highly reflected are now deflected in most cases. When the particles are reflected, they can now reach $72\%$ of light speed. We note that positive slippage also comes with a reduction in the bubble strength and the associated cost in maintaining the negative mass-energy field.

Unfortunately, speeds of particles near the ship can now reach $14\%$ of light speed, a slightly greater hazard than without positive slippage.

So far, we have considered the possibility of interacting with dust. But what about
something larger? It is hypothesized that trillions of rogue planets, i.e. planets that are not gravitationally bound to a star, exist in the
Milky Way~\cite{powell2024}. If a ship is traveling at any significant speed, encountering such an object would result in swift destruction. Detecting and evading
such objects at speed is also likely to be challenging. What, if anything, can be done to guard against this possibility?

\section{Negative Slippage: Dealing with Rogue Planets and Other Massive Obstacles}
\begin{figure}[ht]
    \centering
    \begin{subfigure}{0.33\textwidth}
    \includegraphics[width=\linewidth]{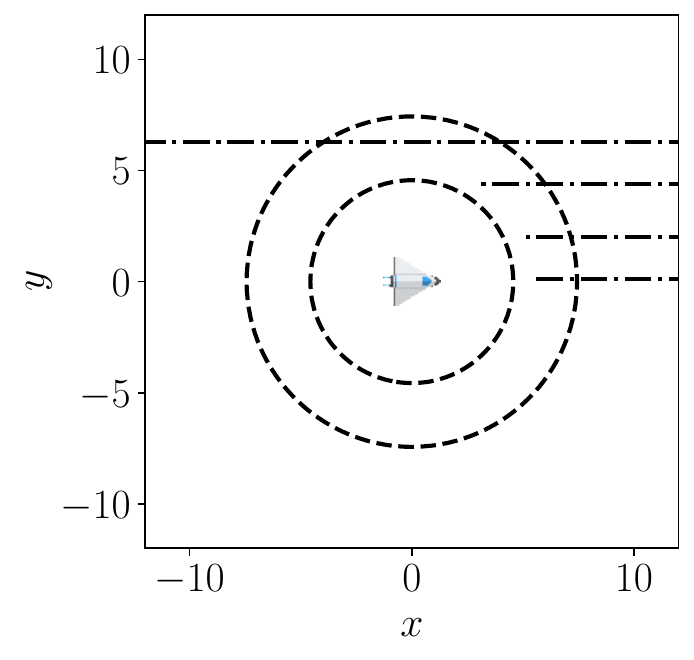}
    \caption{}\label{fig:negative_slippage}
    \end{subfigure}%
    \begin{subfigure}{0.33\textwidth}
    \includegraphics[width=\linewidth]{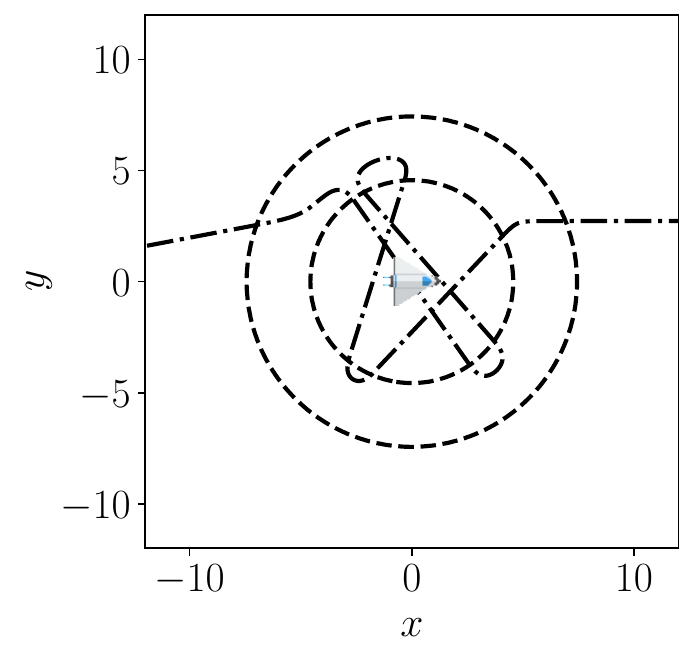}
    \caption{}\label{fig:negative_slippage_left}
    \end{subfigure}%
    \begin{subfigure}{0.33\textwidth}
    \includegraphics[width=\linewidth]{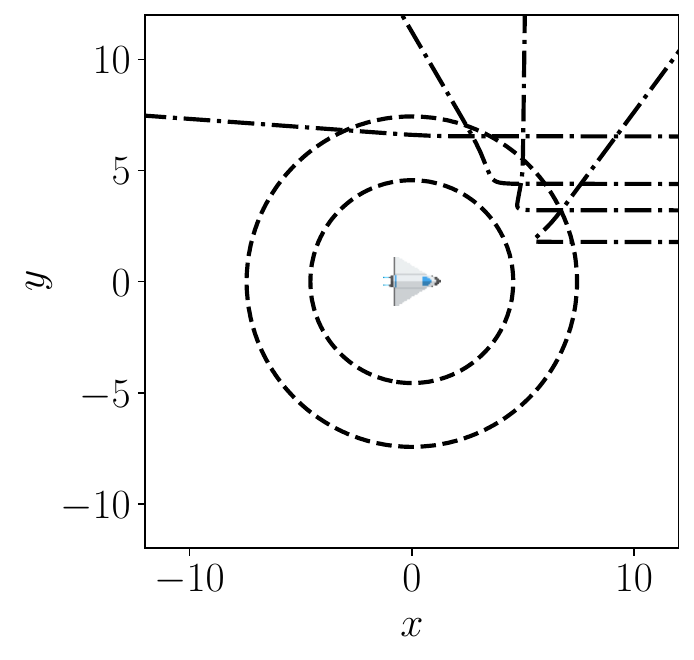}
    \caption{}\label{fig:negative_slippage_right}
    \end{subfigure}%
    \caption{
      Fig~\ref{fig:negative_slippage}: a warp drive moving to the right (positive $x$) through a field of stationary particles with negative slippage. The plot is made in frame co-moving with the ship. The two dashed concentric circles represent the inner and outer radii of the warp field. As without slippage, stationary particles are stopped before they enter the bubble. Fig~\ref{fig:negative_slippage_left} shows a single particle trajectory and makes the point that trajectories can now be quite complicated. Fig.~\ref{fig:negative_slippage_right}: a warp drive moving through a field of particles moving to the right at $1\%$ light speed. The particles are reflected, much as they were without slippage.
    }
    \label{fig:negative_slippage_all}
\end{figure}

Let us consider a ship traveling in a warp bubble, that encounters a large mass, e.g., a rogue planet. Let us further hypothesize that in this case, the warp field would shut down suddenly.
Mathematically, ``shutting-down'' simply means setting $u_d = 0$ in Eq.~\eqref{eq:slippage}. In this scenario, the ship would then suddenly find itself close to a deadly obstacle, traveling at speed $u_s$, and in danger of crashing.

To remedy this, let us then consider the case where $u_d > u_b$ (which we call a \textit{negative slippage}). In this configuration, to remain centered in the warp field, a ship would have to have negative $x$ momentum, i.e. it would be traveling backward while still being dragged forward. This, in turn, implies that once the warp field shuts down, the ship would suddenly reverse its direction, backing away from the obstacle, possibly without the crew feeling acceleration. Precise claims regarding what kind of ``forces'' a crew inside the ship would be subjected to, would require us to model the evolution of the spacetime and equation of state of the warp field. This is beyond the scope of this paper. The work of Clough et al.~\cite{clough2024no}, however, suggests that a collapsing warp bubble might well be dangerous to the crew.

In Fig.~\ref{fig:negative_slippage_all} we see the behavior of a ship passing through a debris field with negative slippage. With stationary particles, Fig.~\ref{fig:negative_slippage}, we see behavior that is very similar to what we saw without slippage. Particles come to a stop before reaching the inner bubble.

In the case of the left-moving particle, Fig.~\ref{fig:negative_slippage_left}, it is now possible to obtain quite complex trajectories. Because of this, we have shown only a single example. We note also that particles with up to $23\%$ of light speed can now be found in the vicinity of the ship. So while this option may theoretically provide some safety against collision with rogue planets, it offers less protection from dust and meteors.

For the right-moving particles, Fig.~\ref{fig:negative_slippage_right}, the situation is no worse than for the original warp drive. Particles are reflected with up to $80\%$ of light speed.

\section{The Deflector Shield}\label{sec:deflector_shild}

To overcome the hazards from all sources considered so far, we introduce an additional term into the warp metric, one that frame drags particles away from the direction of travel.

We now define
\begin{align}
    v^x(t, x, y, z) & = u_d \theta(r;\; R,\, \sigma) \label{eq:shift_vectors_vx}, \\
    v^y(t, x, y, z) & = k\, \rho_y(y,z) \phi(r;\; R,\, {\sigma}) \label{eq:shift_vectors_vy}, \\
    v^z(t, x, y, z) & = k\, \rho_z(y,z) \phi(r;\; R,\, \sigma)\label{eq:shift_vectors_vz},
\end{align}
where
\begin{align}
    \rho_y(y,z) & = \frac{y}{\sqrt{y^2 + z^2 + \epsilon}}, \\
    \rho_z(y,z) & = \frac{z}{\sqrt{y^2 + z^2 + \epsilon}},
    \label{eq:def_rho}
\end{align}
and $\phi(r;\; R,\, \sigma)$ provides a region that is positive inside a spherical shell centered on $R+\sigma$, but which fades radially in both directions across a distance $\sigma$.

We now define $\phi(r;\; R,\, \sigma)$ explicitly as
\begin{equation}
    \phi(r;\; R,\, \sigma) = \theta\left[r - (R + \sigma);\; 0,\, \sigma\right]\,\theta\left[(R + \sigma) - r;\; 0,\, \sigma\right].
    \label{eq:phi_trans_def}
\end{equation}
To paraphrase Alcubierre~\cite{Alcubierre_1994}, a protection mechanism based on such a local distortion of spacetime just begs to be given the familiar name of the ``deflector shield'' of science fiction.

\begin{figure}[h]
    \centering
    \includegraphics[width=0.4\textwidth]{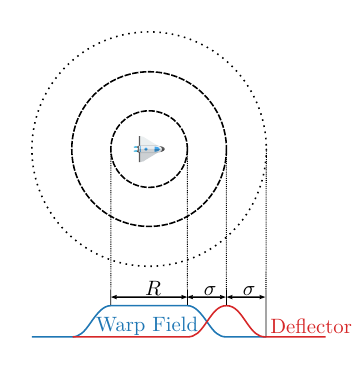}
    \caption{The shape of the deflector shield. The two dashed concentric circles represent the inner and outer radii of the warp field. The two dotted concentric circles represent the radii of the deflection field \rev{(only one dotted circle is visible in the figure, as the inner dotted circle coincides with the warp field's inner dashed circle)}. The innermost edge in the figure is at $r=R$, the next shell is at $r=R+\sigma$, and the final shell is at $r=R+2\sigma$. The deflector begins to turn on at $r=R$ and shuts off at $r=R+2 \sigma$.}
    \label{fig:deflectorshape}
\end{figure}

In Fig.~\ref{fig:deflectorshape}, we show a schematic representation of a ship's warp field combined with this new addition.

We would like to stress that the choice of flow vector made in Eqs.~\eqref{eq:shift_vectors_vx}-\eqref{eq:shift_vectors_vz} still represent a warp drive spacetime of the Nat\'ario class. This means that the same analysis and conclusions of Ref.~\cite{santiago_2022} apply to our deflector shields as well. More specifically, the Stress-Energy-Momentum tensor violates all the energy conditions.

Together with $k$, which we call the \textit{deflection strength}, this new addition causes metric shift vector to repel incoming particles around the warp bubble.

Initially, we will consider a deflection strength of $k=0.9$. This means that particles will be frame-dragged at $90\%$ the speed of light. This is, admittedly, extreme. More so than it probably needs to be, but it reveals an interesting dynamic.

\begin{figure}[ht]
    \centering
    \begin{subfigure}{0.33\textwidth}
    \includegraphics[width=\linewidth]{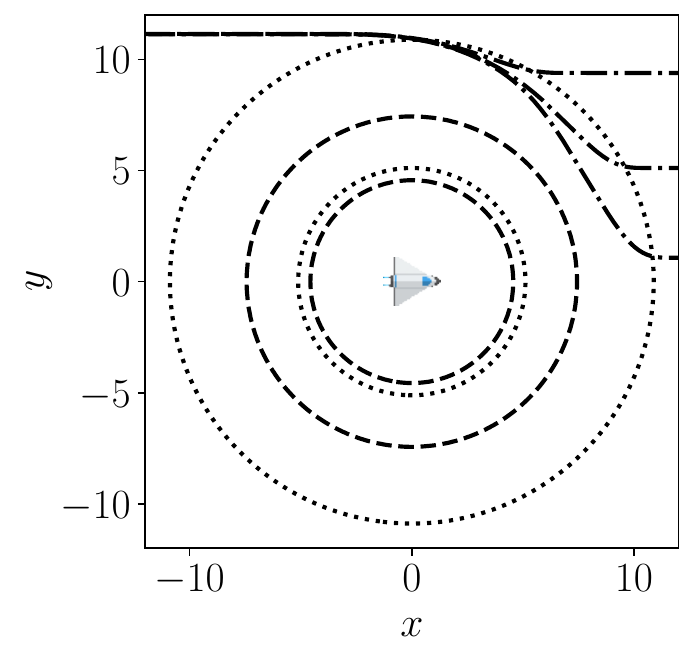}
    \caption{}\label{fig:deflector}
    \end{subfigure}%
    \begin{subfigure}{0.33\textwidth}
    \includegraphics[width=\linewidth]{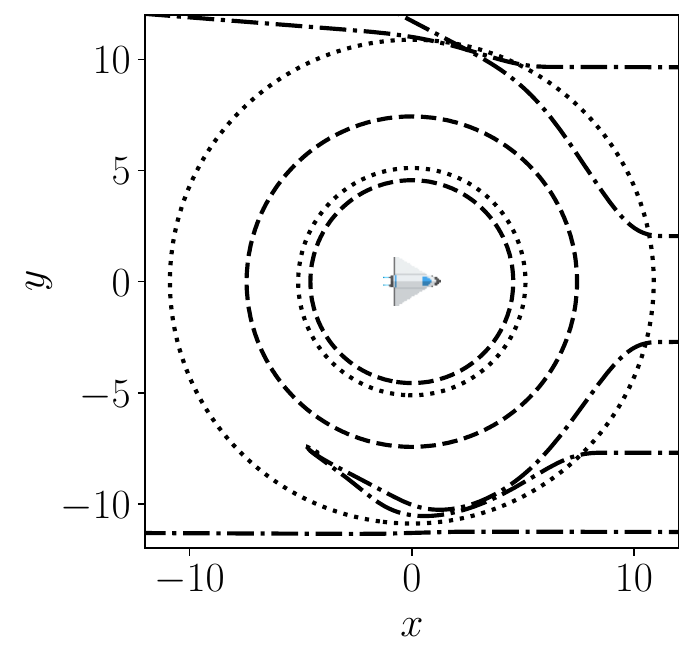}
    \caption{}\label{fig:deflector_left}
    \end{subfigure}%
    \begin{subfigure}{0.33\textwidth}
    \includegraphics[width=\linewidth]{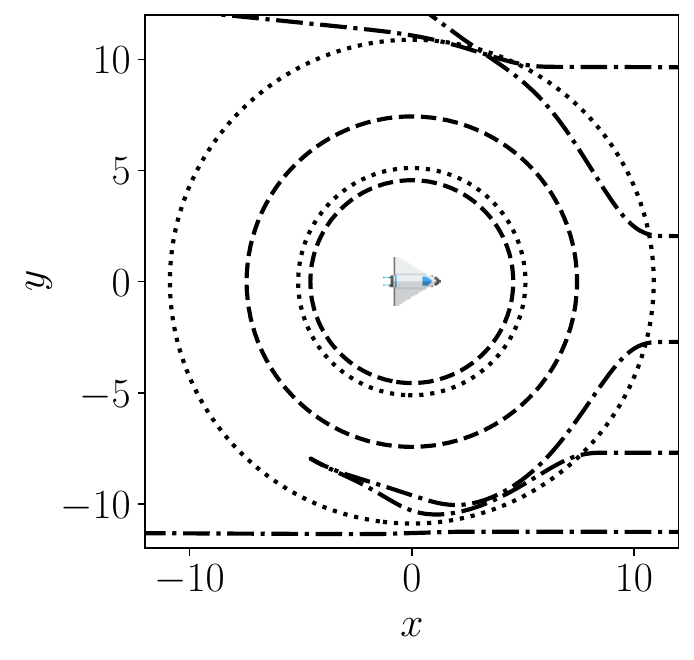}
    \caption{}\label{fig:deflector_right}
    \end{subfigure}%
    \caption{
      A warp drive ship moving to the right through a field of particles with a deflector. The plots are made in a frame co-moving with the ship. The two dashed concentric circles represent the inner and outer radii of the warp field. The two dotted concentric circles represent the radii of the deflection field. In Fig.~\ref{fig:deflector}, we show stationary particles being smoothly swept aside. In Figs.~\ref{fig:deflector_left} and \ref{fig:deflector_right} we show, respectively, left and right moving particles, which are also swept aside. Note, however, that some fraction of them converge to a point that co-moves with the ship.
    }
\end{figure}

With the stationary particles, Fig.~\ref{fig:deflector}, everything seems to be functioning as expected. Particles are swept off to the side and out of the way, never coming near the ship.

However, when the particles are moving to the left (Fig.~\ref{fig:deflector_left}) or right (Fig.~\ref{fig:deflector_right}), we see something unusual. Note that for this particular analysis, we have chosen to impart a very small $y$ component to the particle velocity $V^i=(\pm 0.01, .001, 0)$ because it reveals an interesting behavior.

In both the left and right figures, two of the trajectories converge to a point and stop. Particles that get near this point are usually attracted to it, and those that reach this point accelerate without limit but stay in place. Even photons get stuck at this point. If one chooses $V^i=(\pm 0.1, -.001, 0)$ one finds a symmetric point on the top of the figure. In three dimensions, this region is actually ring-shaped.

What is happening at this point? All the particles in the ring are attempting to converge toward the axis, but are getting frame-dragged onto a straight path. We can find the location of this point in the $xy$-plane from the geodesic equations directly. To do this, we write the full geodesic equations allowing for non-zero $y$ and $V^y$ as
\begin{align}
    \frac{\ud{X}}{\ud{t}} & = u, \label{eq:fixed_point_start} \\
    \frac{\ud{Y}}{\ud{t}} & = V^y + k_0\, \text{sign}\left(Y\right) \, \phi, \label{eq:fixed_point_a}\\
    \frac{\ud{V^x}}{\ud{t}} & = \frac{k_0 V^y \left[\left(u^2 - 1\right)X + u Y V^y\right]\,\text{sign}(Y)\,\phi^\prime}{\sqrt{X^2 + Y^2}},\label{eq:fixed_point_b}\\
    \frac{\ud{V^y}}{\ud{t}} & = \frac{k_0 V^y \left[u X V^y + \left((V^y)^2 - 1\right)Y\right]\,\text{sign}(Y)\,\phi^\prime}{\sqrt{X^2 + Y^2}}.\label{eq:fixed_point_c}
\end{align}
%
For brevity, we've written
\begin{equation}
    \phi = \phi(r;\; R,\, \sigma), \quad \phi^\prime = \frac{\ud{}}{\ud{r}}\phi(r;\; R,\, \sigma).
\end{equation}

Thus, the ``co-moving'' points we seek are given by the $(X, Y, V^y)$ values which make the right-hand sides of Eqs.~\eqref{eq:fixed_point_a}-\eqref{eq:fixed_point_c} identically zero. To find them, we solve Eqs.~\eqref{eq:fixed_point_start}-\eqref{eq:fixed_point_c} by first introducing cylindrical polar coordinates $(r,\vartheta,z)$ centered on the ship, with $x = r \cos{\vartheta}$ and $y = r \sin{\vartheta}$. Next, we use Eq.~\eqref{eq:def_fa} as the form function. Finally, we solve for the regime where the particle is in the transition region, and obtain
\begin{align}
    V^y &=  -\text{sign}\left(Y\right) \sqrt{1-u^2}, \\
    r &= -\frac{\sigma  \sqrt{1-u^2}}{k_0}+R+2 \sigma,\\
    \vartheta &= \tan ^{-1}\left(-u,\sqrt{1-u^2}\right),\\
    X &= -u \left(-\frac{\sigma  \sqrt{1-u^2}}{k_0}+R+2 \sigma \right),\\
    Y &= \pm \left( \frac{\sigma  \left(u^2-1\right)}{k_0}+\sqrt{1-u^2} (R+2 \sigma ) \right).
\end{align}
Mathematically, there are eight roots in the $xy$-plane, two in front and two in back for both the inner region of the deflector, i.e. $R < r < R+\sigma$ and for the outer region $R+\sigma < r < R+2 \sigma$. In addition, any particle placed at a point with $\phi'=0$ can co-move with the ship. However, only the two trailing points identified above seem to act as attractors. When they appear, these co-moving points continuously accelerate massive particles or blue-shift photons, and probably represent a significant energy cost for the ship.

\begin{figure}[ht]
    \centering
    \begin{subfigure}{0.4\textwidth}
    \includegraphics[width=\linewidth]{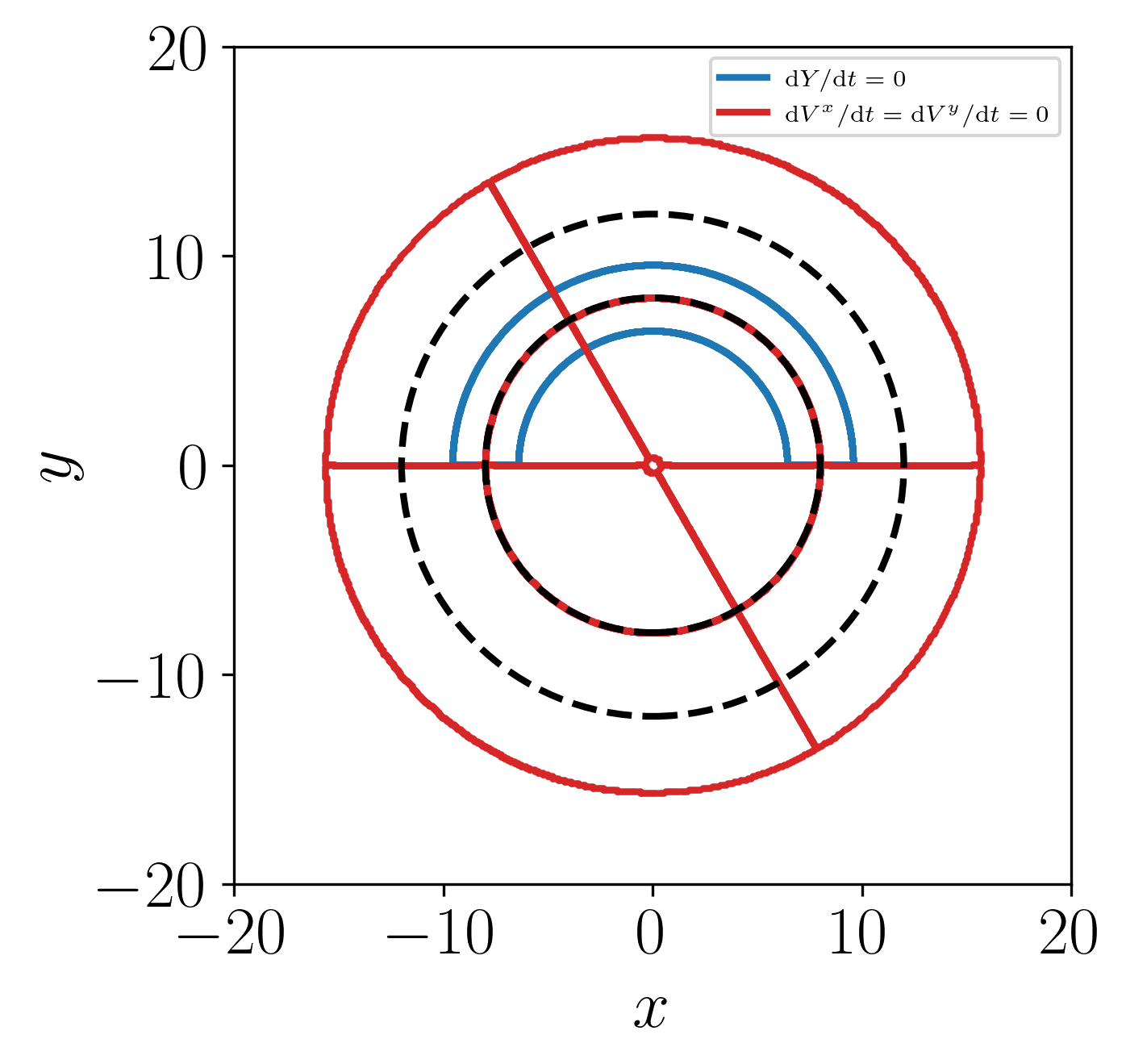}
    \caption{}\label{fig:fixed_point_0}
    \end{subfigure}%
    \begin{subfigure}{0.4\textwidth}
    \includegraphics[width=\linewidth]{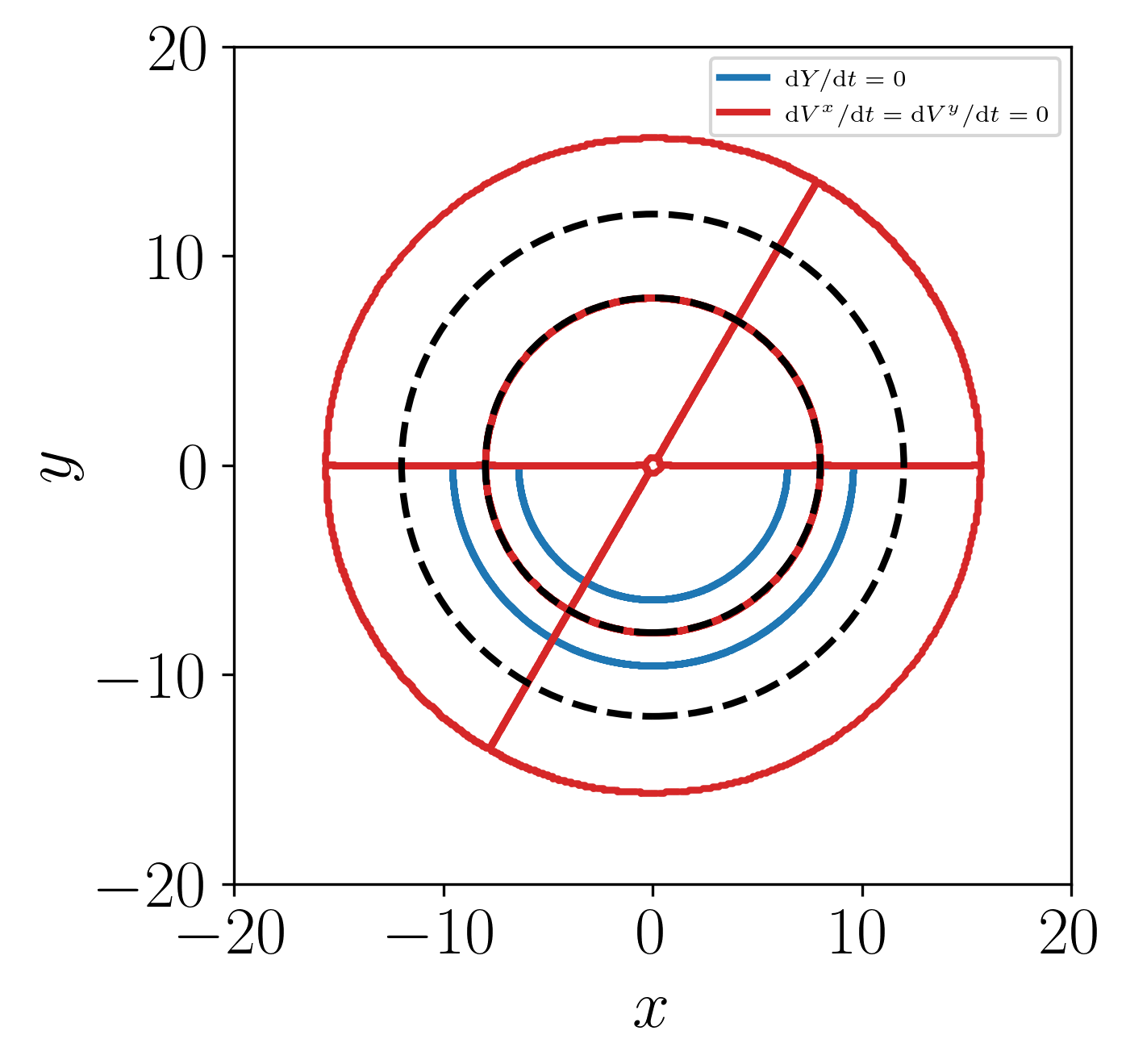}
    \caption{}\label{fig:fixed_point_1}
    \end{subfigure}%
    \caption{Co-moving points of the system depicted in Fig.~\ref{fig:deflector_left}. In panel (a) and (b), co-moving points for $V^y = \pm \sqrt{1-(1/2)^2}$ are depicted, respectively. 
    }
    \label{fig:fixed_point_all}
\end{figure}

Alternatively, we can see these co-moving points by looking at the intersection of the zero contour levels of the right-hand sides of Eqs.~\eqref{eq:fixed_point_a}-\eqref{eq:fixed_point_c}. We represent these contour levels by different colors in Fig.~\ref{fig:fixed_point_all} for the configuration depicted in Fig.~\ref{fig:deflector_left}. We depict the deflection radii ($R + \sigma$ and $R+ 2\sigma$) in black dashed lines. In order to create a level plot, we need to choose a value of $V^y$, and for this we always choose $\sqrt{1-u^2}$, because that is the speed required for a point to co-move.

Another thing our solution above tells us about these co-moving points is that they only occur when the deflector strength is sufficiently large, i.e. when $k_0 > \sqrt{1-u^2}$. If one wishes to avoid these co-moving points, this would constrain the deflector strength at higher bubble speeds. For a ship traveling at $u=1/2$, this means the deflector strength should be kept below $k=\sqrt{3}/2$.

However, even though keeping the deflector strength below this critical value prevents the co-moving points from occurring, the wake of the ship is still plagued with high accelerations and complex particle dynamics.

\begin{figure}[ht]
    \centering
    \begin{subfigure}{0.33\textwidth}
    \includegraphics[width=\linewidth]{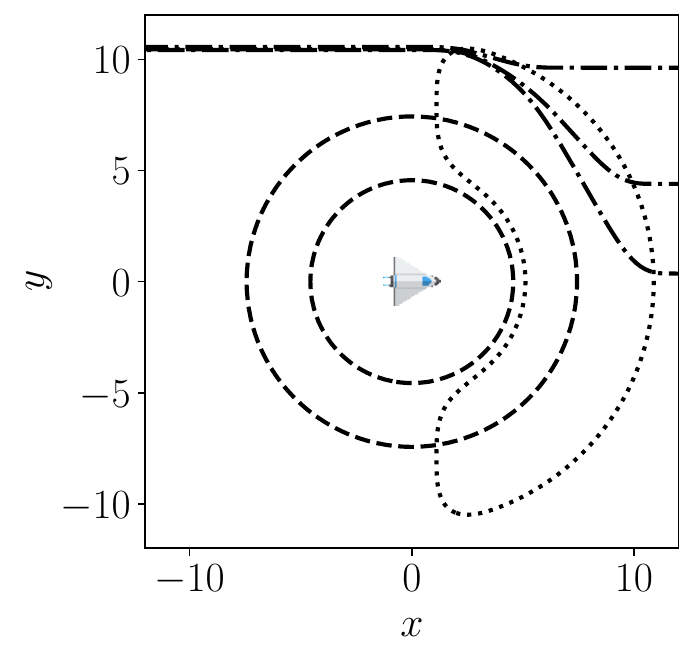}
    \caption{}\label{fig:deflector_noback}
    \end{subfigure}%
    \begin{subfigure}{0.33\textwidth}
    \includegraphics[width=\linewidth]{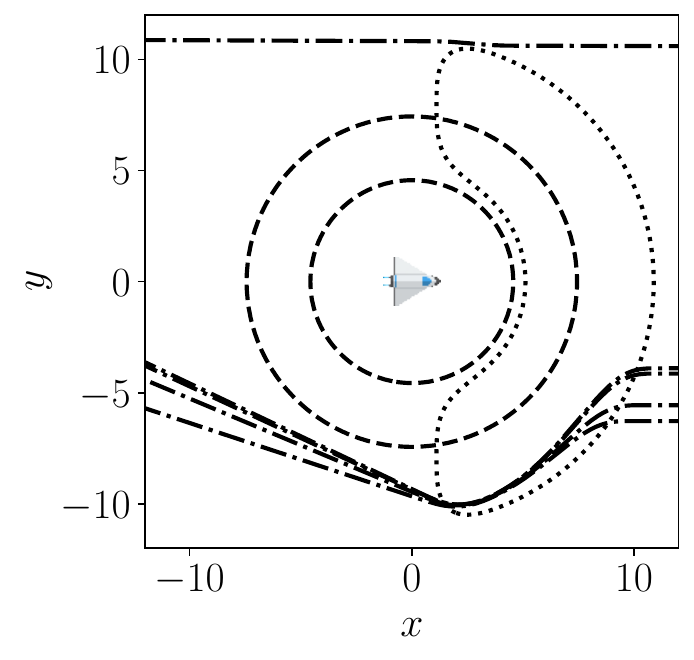}
    \caption{}\label{fig:deflector_noback_left}
    \end{subfigure}%
    \begin{subfigure}{0.33\textwidth}
    \includegraphics[width=\linewidth]{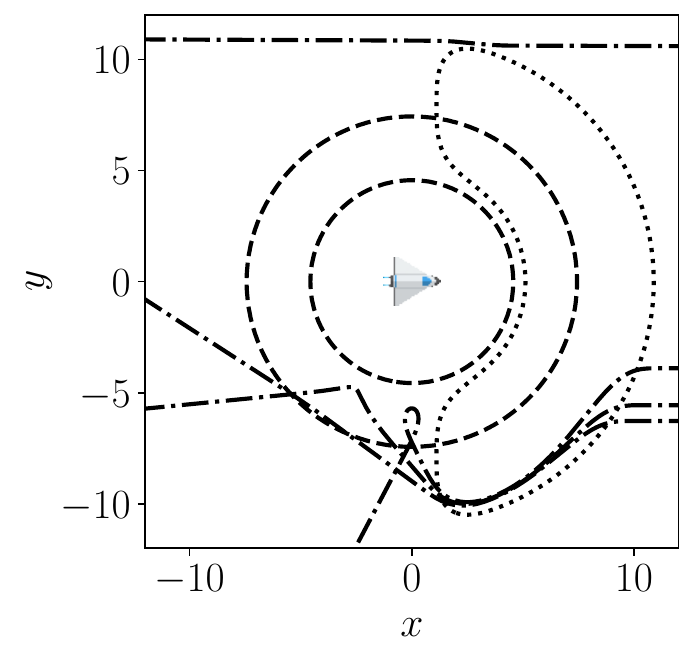}
    \caption{}\label{fig:deflector_noback_right}
    \end{subfigure}%
    \caption{
      A warp drive ship moving to the right through a field of particles with the back portion of the deflector turned off. The plots are made in a frame co-moving with the ship. The two dashed concentric circles represent the inner and outer radii of the warp field. The dotted curve represents the 99\% strength contour level for the deflection field. In Figs.~\ref{fig:deflector_noback}, \ref{fig:deflector_noback_left} and, \ref{fig:deflector_noback_right}, we show the deflection of stationary, left and right moving particles, respectively. Note that, in this ``half-deflector'' configuration, we avoid the co-moving points represented in Figs.~\ref{fig:deflector_left} and \ref{fig:deflector_right}.
    }
    \label{fig:deflector_noback_all}
\end{figure}

To compensate for this effect, we turn off the deflector shield in the rear of the ship by slightly modifying Eq.~\eqref{eq:phi_trans_def} to be
\begin{multline}
    \phi(r;\; R,\, \sigma) = \theta\left[r - (R + \sigma);\; 0,\, \sigma\right] \theta\left[(R + \sigma) - r;\; 0,\, \sigma\right] \times \\ 
    \left\{(1-B) \, \theta\left[(R + \sigma) - x;\; 0,\, \sigma\right] + B\right\}.
    \label{eq:phi_trans_def_mod}
\end{multline}
The new parameter $B$, which we refer to as \textit{deflector back}, accomplishes this. By making $B=0$, we shut down the deflection field on the rear half of the warp bubble. By making $B = 1$, however, we restore a deflection field that encompasses the whole warp field.

The results of this modification can be seen in Fig.~\ref{fig:deflector_noback_all}. The problematic ring-shaped region is now gone. While we do have some complex trajectories, particles never attain speeds greater than $52\%$ of light speed and so there should be less strain on the ship's power systems.

\begin{figure}[ht]
    \centering
    \begin{subfigure}{0.33\textwidth}
    \includegraphics[width=\linewidth]{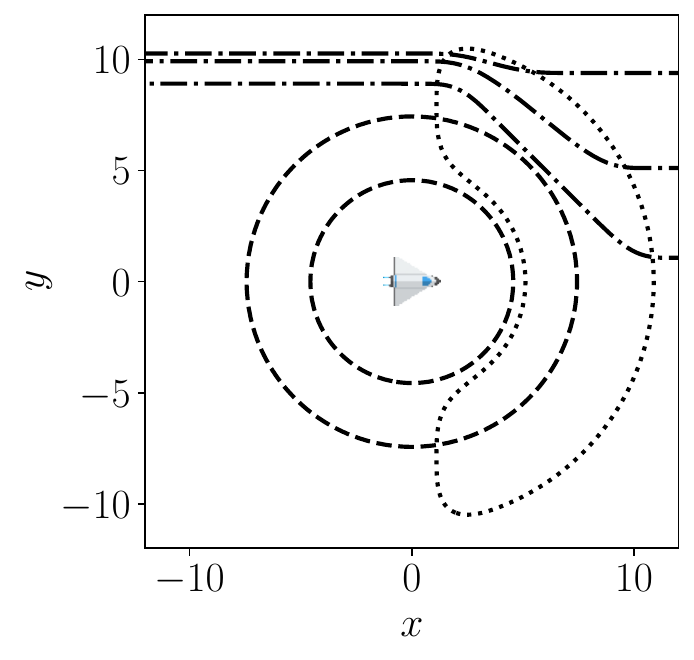}
    \caption{}\label{fig:optimal}
    \end{subfigure}%
    \begin{subfigure}{0.33\textwidth}
    \includegraphics[width=\linewidth]{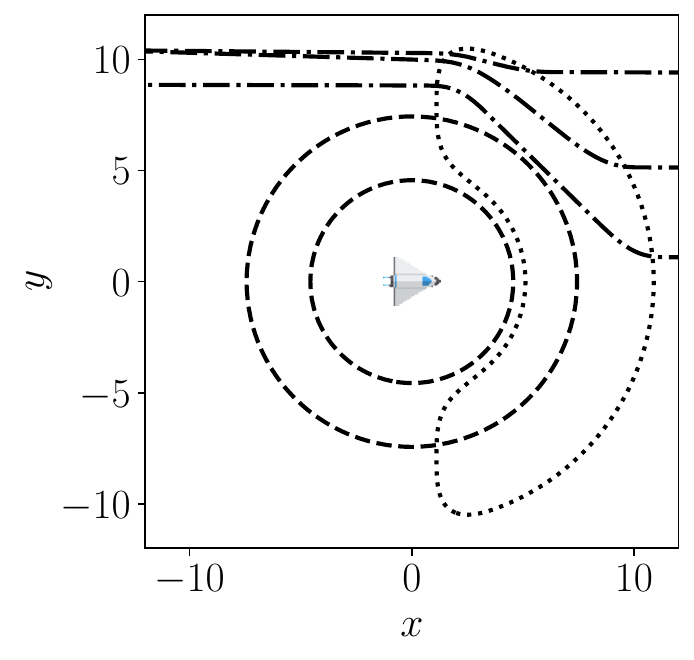}
    \caption{}\label{fig:optimal_left}
    \end{subfigure}%
    \begin{subfigure}{0.33\textwidth}
    \includegraphics[width=\linewidth]{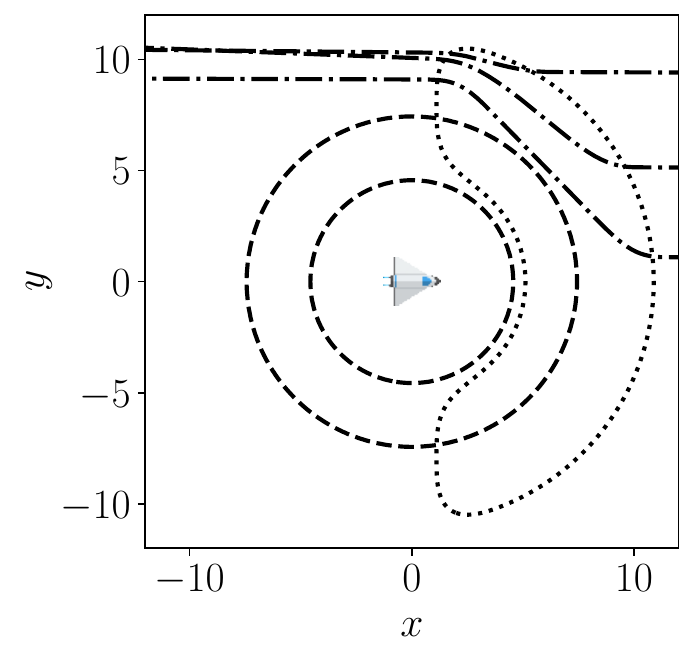}
    \caption{}\label{fig:optimal_right}
    \end{subfigure}%
    \caption{
      An ``optimal'' configuration of the warp drive plus deflector shield system. Once again, we show a warp drive ship moving to the right through a field of particles with the back portion of the deflector turned off. The plots are made in a frame co-moving with the ship. The two dashed concentric circles represent the inner and outer radii of the warp field. The dotted curve represents the 99\% strength contour level for the deflection field. Fig~\ref{fig:optimal} shows the behavior for static particles, Fig~\ref{fig:optimal_left} shows the behavior for left-moving particles, and Fig.~\ref{fig:optimal_right} right-moving particles.
    }
    \label{fig:optimal_all}
\end{figure}

Finally, putting together all the elements of the discussion, let us consider a gentler deflector shield ($k=0.45$ and turned off in the back) combined with a modest negative slippage ($u_s=-0.01$). Using this combination, the ship can safely navigate a field of particles with speeds of $1\%$ light speed or less, moving in random directions. In the event of warp field shut down, the ship will move backwards at $1\%$ light speed. We show the results of such a configuration in Fig.~\ref{fig:optimal_all}. There is very little difference in the behavior of the three cases, no complex trajectories, and no particles getting close to the ship. The fastest moving particle in any of these simulations is just a little over $2\%$ light speed. We will call this our ``optimal'' warp drive plus deflector shield combination.

\subsection{Particle deflection on the Natário drive}

In constructing our deflector shield modification, we have elected to work with Alcubierre's framework because it is mathematically simpler, and because it allows us to more readily decouple and tune the deflector effect from the warp drive itself.

However, we have noted that the zero expansion warp drive described By Nat\'ario in Ref.~\cite{Natario_2002}\footnote{This refers to a particular spacetime, which, perhaps confusingly, is also part of the broader Nat\'ario \textit{class} of spacetimes.}, because of the way it prevents expansion and contraction of the spacetime metric, inherently has a deflector shield effect, which we will further explore in this section.

The components of the flow vector of the Nat\'ario drive were given explicitly in Sec. 2 of Ref.~\cite{Natario_2002} in spherical coordinates. To allow integration with our simulator code, we have transformed these components back to cartesian coordinates and obtained, 
\begin{align}
    v^x(t, x, y, z) & = (u/2) \left(2 \theta +(y^2+z^2) / r \, \theta^\prime\right) \label{eq:natario_shift_vectors_vx}, \\
    v^y(t, x, y, z) & = -(u/2)((x-u\,t)\,y) / r \, \theta^\prime \label{eq:natario_shift_vectors_vy}, \\
    v^z(t, x, y, z) & = -(u/2)((x-u\,t)\,z) / r \, \theta^\prime \label{eq:natario_shift_vectors_vz},
\end{align}
where $u$ denotes the bubble's speed, $r$ is the distance function given by Eq.~\ref{eq:alcubierre_r}, $\theta$ denotes the form function of Eq.~\eqref{eq:trans_poly}, and, $\theta^\prime$ its derivative (with respect to $r$).

Its easy to verify that with $v^i$ given by Eqs.~\eqref{eq:natario_shift_vectors_vx}--\eqref{eq:natario_shift_vectors_vz}, we have
\begin{equation}
  \partial_x v^x + \partial_y v^y + \partial_z v^z = 0.
  \label{eq:natario_divergence}
\end{equation}

\begin{figure}[ht]
    \centering
    \begin{subfigure}{0.33\textwidth}
    \includegraphics[width=\linewidth]{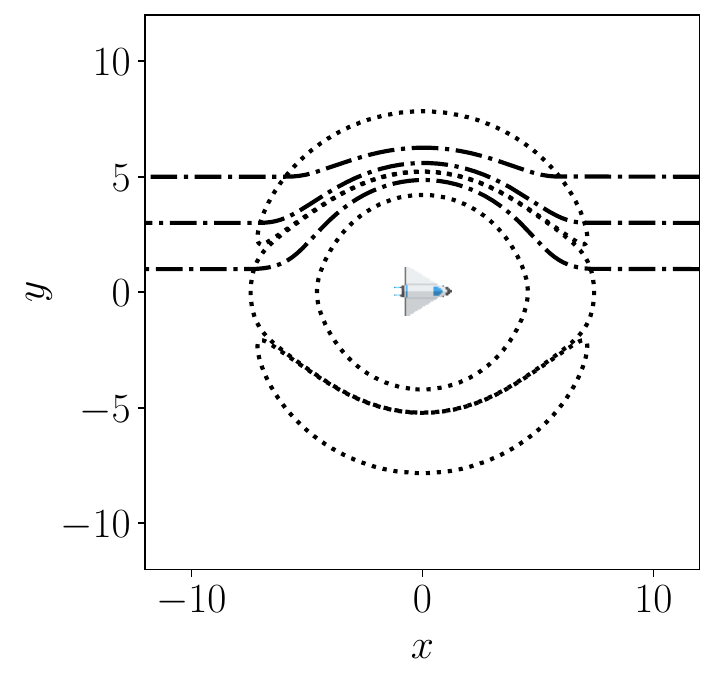}
    \caption{}\label{fig:natario}
    \end{subfigure}%
    \begin{subfigure}{0.33\textwidth}
    \includegraphics[width=\linewidth]{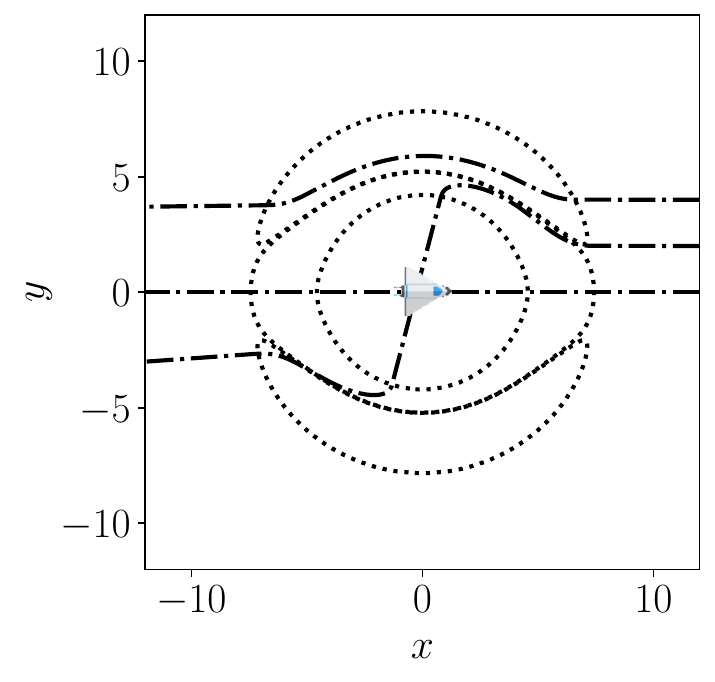}
    \caption{}\label{fig:natario_left}
    \end{subfigure}%
    \begin{subfigure}{0.33\textwidth}
    \includegraphics[width=\linewidth]{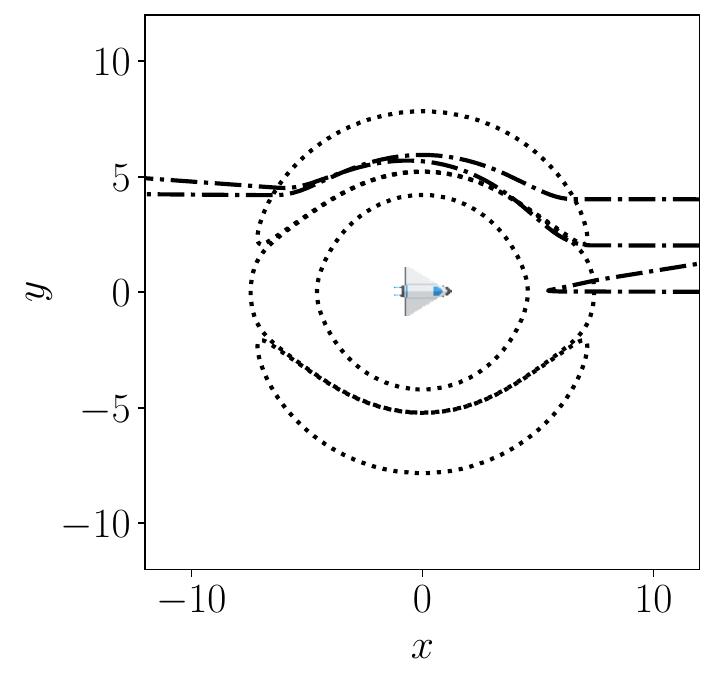}
    \caption{}\label{fig:natario_right}
    \end{subfigure}%
    \caption{
      Fig~\ref{fig:natario}: a Natário warp drive ship moving to the right (positive $x$) through a field of stationary particles. The two dotted concentric ovoids (the ``eye'') represent the inner and outer radii of the warp field where it is positive. The ``eyelid'' regions are where the warp drive field is negative. Particles are deflected around the ship. Fig.~\ref{fig:natario_left} Particles moving to the left with a speed of $-.01 c$ that impact the warp bubble are focused through the center, achieving speeds of $.1$ relative to the ship and thus presenting a significant hazard. Fig.~\ref{fig:natario_right}: Particles are moving to the right at $1\%$ light speed. Particles are reflected by the warp field in a very small region at the front, and reach $80\%$ light speed. The rest flow around the ship.
    }
    \label{fig:natario_warp_all}
\end{figure}

\section{Conclusion}

In this work, we have explored the interactions of particles with an Alcubierre Drive, focusing on safety and energy considerations when interacting with dust, debris, and rogue planets.

We have shown that the Alcubierre and Natário Drives offer
some protection against these hazards, but not enough to make travel safe. We have also shown that it is possible, with small modifications of the field, to deflect incident particles away from a ship using an Alcubierre drive without accelerating them unduly. We have noted, along the way, some surprisingly complex trajectories that can arise from such straightforward modifications.

As an aid to understanding the effects of the Alcubierre Drive on interstellar hazards, we provide an open-source, interactive visualization tool written in \texttt{Rust}. Using this tool, it is possible to explore the surprisingly complex dynamics of particles interacting with warp drives and deflector shields. Additionally, with the code, it is possible to push the deflector shield out further and modify its thickness independently of the thickness of the warp bubble.

We note that the parameter space for this problem is huge and that in this work we have barely scratched the surface of phenomena.

\section*{Acknowledgments}

We would kindly like to thank Erik Schnetter, Helvi Witek, Peter Diener, Roland Hass
 and Zachariah B. Etienne for helpful comments, discussions and suggestions on improving this work. LTS would also like to thank Jessica Santiago for answering questions and providing relevant references regarding warp drive research during the elaboration of this work.

This research was inspired by Katy Clough's talks about warp drive simulations given at the Einstein Toolkit workshops in Aveiro, Portugal (2023), and Baton Rouge, Louisiana USA (2024).

The spaceship artwork used in this paper is licensed under the \texttt{CC-BY 4.0} license and was obtained from Ref.~\cite{spaceships}.

In addition, we gratefully acknowledge NSF grant 2411068.


\section*{References}
\bibliographystyle{unsrt}
\bibliography{ref}
\end{document}